# A Model of the Normal State Susceptibility and Transport Properties of Ba(Fe$_{1-x}$Co$_x$)$_2$As$_2$: Why does the magnetic susceptibility increase with temperature?

B.C. Sales, M. A. McGuire, A. S. Sefat and D. Mandrus
*Materials Science and Technology Division, Oak Ridge National Laboratory*
*Oak Ridge, TN 37831*

**Abstract**

A simple two band model is used to describe the magnitude and temperature dependence of the magnetic susceptibility, Hall coefficient and Seebeck data from undoped and Co doped BaFe$_2$As$_2$. Overlapping, rigid parabolic electron and hole bands are considered as a model of the electronic structure of the FeAs-based semimetals. The model has only three parameters: the electron and hole effective masses and the position of the valence band maximum with respect to the conduction band minimum. The number of extrinsic electrons at T=0 are estimated from experiment. The model is able to reproduce in a semiquantitative fashion the magnitude and temperature dependence of many of the normal state magnetic and transport data from the FeAs-type materials, including the ubiquitous increase in the magnetic susceptibility with increasing temperature.

**Introduction**

The new Fe based superconductors [1-8] with transition temperatures as high as 55 K have attracted considerable interest within the condensed matter physics community. Two unusual normal state properties exhibited by all of the different families of FeAs based superconductors (LaFeAsO, BaFe$_2$As$_2$, LiFeAs) are a magnetic susceptibility that increases more or less linearly with temperature [1,9-14] up to at least 700 K [15] and a Seebeck coefficient that is large ($\approx$ 50-90 µV/K) and often exhibits a maximum near 100 K.[9,11,16] State-of-the art electronic structure calculations indicate small compensating electron and hole Fermi surfaces with a high density of states.[17-19] The Fermi surface consists of three hole sheets and two electron sheets, with the hole sheets derived from heavier bands (lower velocity). As will be illustrated below, however, many of the normal state properties can be understood by considering a semimetal with one electron and one hole band. The model will be applied to one of the best characterized of the FeAs type superconductors the Ba(Fe$_{1-x}$Co$_x$)$_2$As$_2$ system [12-15,20] where Co doping adds electrons to the Fe d bands near the Fermi energy in a nearly linear fashion. [21] This system has also been studied with most modern condensed matter physics experimental techniques such as elastic and inelastic neutron scattering (INS) [22,23], angle-resolved photoemission spectroscopy (ARPES)[24,25] and atomic resolution scanning tunneling spectroscopy.[26] The simple predictions of a two band model will be compared to some of the normal state properties reported for Ba(Fe$_{1-x}$Co$_x$)$_2$As$_2$ single crystals such as the temperature dependence of the magnetic susceptibility, the Hall coefficient, and the resistivity. The expected variation of the Seebeck coefficient with temperature and doping is calculated from the model.

**The Model**

The essence of the model is shown in Fig. 1. Two parabolic bands are considered with the energy of the electron band given by $E = h^2 k^2/(8\pi^2 m_e^*)$ and the hole band by $E = E_h - h^2(k-k_o)^2/(8\pi^2 m_h^*)$. The bottom of the conduction band, $E_e$, is defined to be zero energy, and the top of the valence band is given by $E_h$. The bands are assumed to be rigid meaning that as electrons are added to the system, $E_e$, $E_h$, $m_e^*$ and $m_h^*$ are constant. In pure BaFe$_2$As$_2$ (above the structural/magnetic transition) the crystals are n-type and the extrapolated excess electron concentration at T=0 is about $1.4 \times 10^{21}$ or 0.07 electrons per Fe atom. As x is increased the measured or estimated carrier concentration (at T=0), $N_0$, increases linearly with x [21]. The value of $N_0$ determines the chemical potential at T=0 ($E_F$ in Fig. 1). As the temperature is increased, the chemical potential is determined numerically from the charge balance constraint [27,28] : $N = P + N_0$, where N is the total number of electrons, and P is the number of holes and

$$N = Nc \cdot \frac{2}{\sqrt{\pi}} F_{1/2}(E_f/k_B T) \qquad (1)$$

where, $N_c = 2(2\pi m_e^* k_B T/h^2)^{3/2}$, $\eta_f = E_f/k_B T$ and

$$F_{1/2}(\eta_f) = \int_0^\infty \frac{\eta^{1/2}}{1+e^{(\eta-\eta_f)}} d\eta \qquad (2)$$

and

$$P = N_v \frac{2}{\sqrt{\pi}} F_{1/2}(\frac{E_h - E_f}{k_B T}) \qquad (3)$$

where $N_v = 2(2\pi m_h^* k_B T/h^2)^{3/2}$, $\eta_f = (E_h - E_f)/k_B T$

These equations are standard for semiconductors or semimetals [27,28]. Once the position of $E_f$ is determined at each temperature, the magnetic susceptibility is determined from [29]

$$\chi = -\mu_B^2 \int_0^\infty \frac{df}{dE} g(E) dE , \qquad (4)$$

where f is the Fermi function and g(E) is density of states from both the electron and hole bands which are both proportional to $E^{1/2}$ in this simple model. For large electron doping or if the hole band is moved far away from the electron band, the Pauli susceptibility for a free electron model is recovered ($\chi = 3N \mu_B^2/2E_f$)

For a two band model the Hall number, $R_H = 1/ec \cdot (P-Nb^2)/(P+Nb)^2$, where $b = \mu_e/\mu_h$ is the mobility ratio between electrons and holes.[30] As a crude approximation, we take $b = m_h^*/m_e^*$, which implies that the scattering rates for electrons and holes are the same. The model is not very sensitive to this assumption and a value of $b=1$ also gives reasonable results.

The Seebeck coefficient is given by the standard transport integrals [28] i.e:

$$S_e = \frac{k_B}{e}[\eta_f - \frac{(r+5/2)F_{r+3/2}(\eta_f)}{(r+3/2)F_{r+1/2}(\eta_f)}] \quad (5)$$

$$F_n(\eta_f) = \int_0^\infty \frac{\eta^n}{1+e^{(\eta-\eta_f)}} d\eta \quad (6)$$

This is the Seebeck coefficient from the electron band and the expression for the hole band is similar except $\eta_f = E_f/k_BT$ is replaced by $\eta_f = (E_h-E_f)/k_BT$. The value of r, a parameterization of the energy dependence of the scattering time, is taken to be the standard value of -0.5.[28] The Seebeck coefficient from two bands is the average Seebeck value weighted by electrical conductivity of each band, i.e.

$$S = \frac{(S_e\sigma_e + S_h\sigma_h)}{(\sigma_e + \sigma_h)} \approx \frac{(S_eN/m_e^* + S_hP/m_h^*)}{(N/m_e^* + P/m_h^*)} \quad (7)$$

where we have again approximated that the mobility of the carriers (electrons or holes) is inversely proportional to the effective mass. These equations are also standard and often used to describe transport in thermoelectric materials [28, 31] The carrier concentration, magnetic susceptibility, Hall and Seebeck coefficients do not depend directly on the scattering rate of the carriers. To estimate the electrical conductivity, however, some assumptions must be made about the dominant scattering mechanisms. The experimental results of Rullier-Albenque et al [21] indicate that for Co doped $BaFe_2As_2$, the dominant scattering mechanism up to at least 150 K is electron-electron scattering and the residual scattering rate at low temperatures is independent of Co doping. The simplest expression for the electrical resistivity incorporating these experimental observations is:

$$\rho = \rho_1/N_o + \rho_2T^2/(N/m_e^*+P/m_h^*) \quad (8)$$

where $\rho_1$ and $\rho_2$ are constants that are adjusted using the Rullier-Albenque et al. data[21] for a particular Co doping x and then held constant for all other x.

**Comparison: Model and Experimental Data**

One set of model parameters was used to compare the results of the model with existing experimental data. Because of the simplicity of the model, there was no attempt to obtain the "best fit" for all of the magnetic and transport data. The goal was to see if a simple model could semiquantitatively account for the magnitude and general trends displayed by the experimental data. Such models are often of great use to experimentalists in providing a conceptual picture that captures essential features of the physics.

The values of the parameters used are $m_e^*=17\ m_0$, $m_h^*=30\ m_0$, $E_h = 250$ K. The scattering exponent used for the calculation of the Seebeck coefficient was taken to be the standard value or r=-0.5 (see Eq. 5) [28]. Although the value of $E_h$ is adjusted to give a good description of the data, this value is close to the values estimated from ARPES data [24,25]. The large values for the effective masses within the context of a free electron-like model are consistent with the high density of states at the Fermi energy expected from detailed modern electronic structure calculations.[17-19] and reflect the d-band character of the states near the Fermi energy. A larger hole effective mass is also consistent with electronic structure calculations.

The initial motivation for the construction of this model was an attempt to understand why the high temperature magnetic susceptibility increases approximately linearly with temperature for the all of the Fe-As compounds- even the compounds that exhibit long range magnetic order at lower temperatures. For example, $BaFe_2As_2$ exhibits long range antiferromagnetic order below 130 K, [22] yet above this temperature the susceptibility increases linearly with temperature (Fig.1a) up to at least 700 K [15]. A similar increase in susceptibility is also observed for Cr metal above a spin density wave (SDW) transition near room temperature.[15, Ref. 30 p 437 ] with no evidence of a maximum or Curie-Weiss behavior up to 1700 K. In both the Fe-As compounds and Cr metal the magnetic transition is attributed to the nesting of hole and electron regions of the Fermi surface.

The magnetic susceptibility results from the model (Fig. 1b) are compared to the measured magnetic susceptibility data from several Co-doped $BaFe_2As_2$ crystals (Fig 1a). The magnitude and general slope of the magnetic susceptibility data above the magnetic/structural transition are reproduced well by the model. With increasing x the susceptibility near room temperature decreases by a similar amount for both the measured data and the model results. For much higher values of Co-doping ,x , the model predicts that the susceptibility should become larger and exhibit a much weaker temperature dependence as the Pauli limit for a single electron band is approached ($\chi = 3N\ \mu_B^2/2E_f$).

The variation of the electron, N, and hole, P, concentrations with temperature are illustrated in Fig. 2a for x=0.1. By construction, the hole concentration at T=0 is 0, and the electron concentration at T=0, $N_0$, increases linearly with cobalt doping starting from a value of $N_0 = 1.4 \times 10^{21}$ electrons/cm$^3$ for x=0.[21]. With increasing temperature $N = P + N_0$. The variation of the apparent electron concentration measured in a Hall experiment as a function of temperature and x is shown in Fig 2b. For temperatures above the magnetic/structural transitions, the model results are remarkably similar in shape and magnitude to the Hall data reported by Rullier-Albenque et al, Fig. 2 [21].

Resistivity data depend directly on the electron or hole scattering rates. The analysis of Rullier-Albenque et al [21] indicated that electron-electron scattering was dominant up to a least 150 K. They also pointed out that this did not necessarily imply unusually strong electron-electron correlations since in the simplest expressions the scattering rate is inversely proportional to the Fermi energy [32], which is very small in these compounds. These experimental results naturally suggest the simple expression for the resistivity given in Eq. 8. The results generated by the model using this expression [Fig. 4] are very similar to the resistivity data reported by Rullier-Albenque et al. as well as resistivity data measured by us and several other groups. [12-14, 33]

The Seebeck coefficient versus temperature calculated from the model using the same parameters is shown in Fig. 5 for several values of x. The magnitude and maximum of S near 100-150 K for x <0.1 is similar to that reported for several of the Fe-As materials. [9,11,16]. The general shape of S(T) agrees with experiment and for x>0.03 the model values of S are in fair agreement with the data reported by Mun et al. [34](in their Fig. 3) which appeared after the preparation of this manuscript.

**Conclusions**

Many of the normal state magnetic and transport properties of the Fe-As type superconductors can be understood within the framework of a simple two band model. The magnetic susceptibility of most semimetals should increase with increasing temperature.

**Acknowledgements**

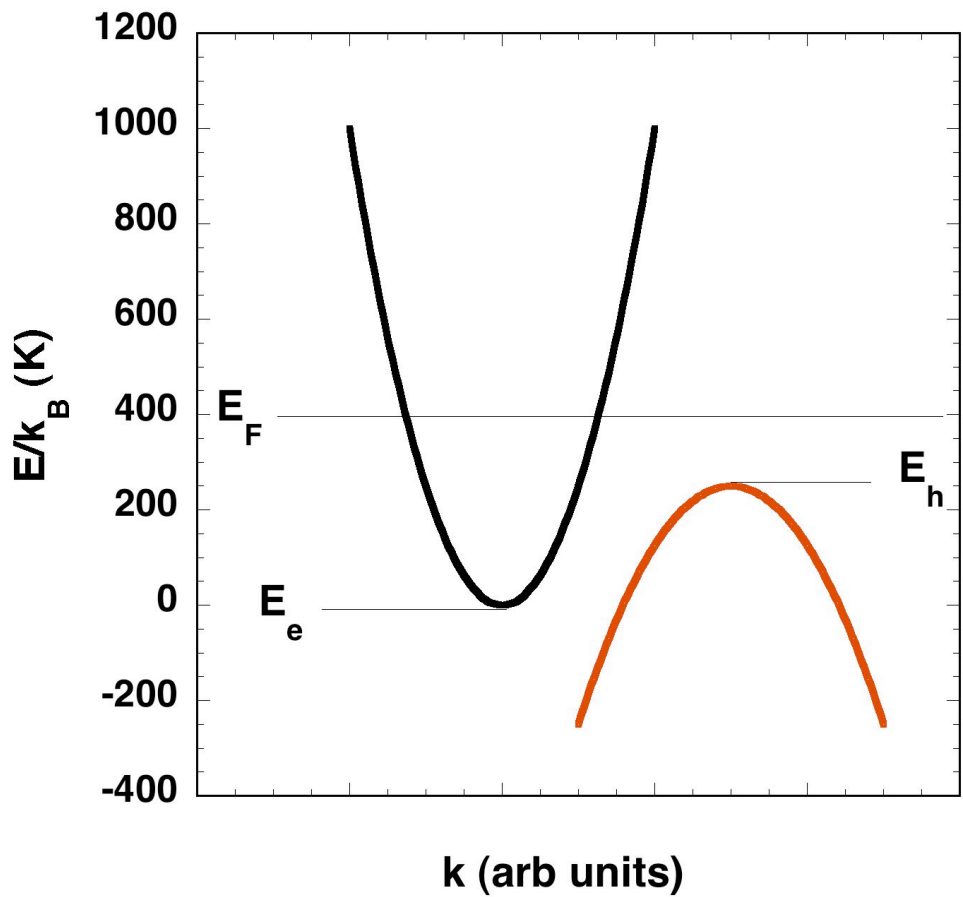

Fig. 1. Schematic of two band semimetal model used to calculate the normal state magnetic susceptibility and transport properties of Ba(Fe$_{1-x}$Co$_x$)$_2$As$_2$ alloys. The electron and hole model bands are assumed parabolic. The conduction band minimum, $E_e$, valence band maximum, $E_h$, and Fermi energy, $E_F$, at T=0 are noted in the figure.

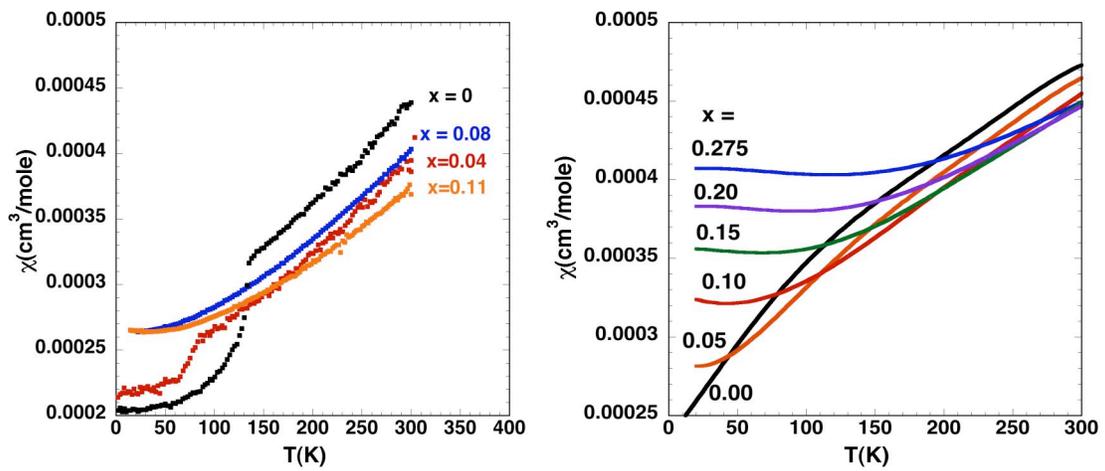

Fig 2. (a) Magnetic susceptibility vs temperature for several $Ba_{0.5}(Fe_{1-x}Co_x)As$ crystals with H= 5T and H perpendicular to *c* (Only data for $T>T_c$ is shown) (b) Calculated susceptibility from the two band model with $m_e^*=17m_0$, $m_h^*=30m_0$, and $E_h/k_B =250$ K.

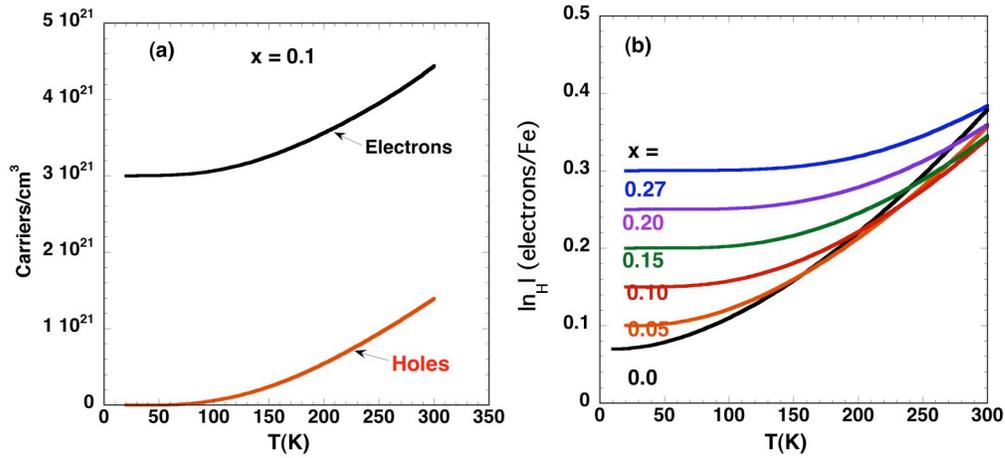

Fig. 3 (a) Calculated temperature dependence of the electron and hole carrier concentration for x = 0.1. (b) Calculated variation of the apparent electron concentration measured in a Hall experiment vs. temperature and x for $Ba(Fe_{1-x}Co_x)_2As_2$. For a two band model the Hall number, $R_H = 1/ec \cdot (P-Nb^2)/(P+Nb)^2$, where $b = \mu_e/\mu_h$ is the mobility ratio between electrons and holes. The model data are in semiquantitative agreement with the Hall data reported by Rullier-Albenque et al. in Fig 2 of Ref. 21 for temperatures above $T_c$ and the magnetic/structural phase transitions. All of the curves were generated with $m_e^* = 17m_0$, $m_h^* = 30m_0$ and $E_h/k_B = 250$ K.

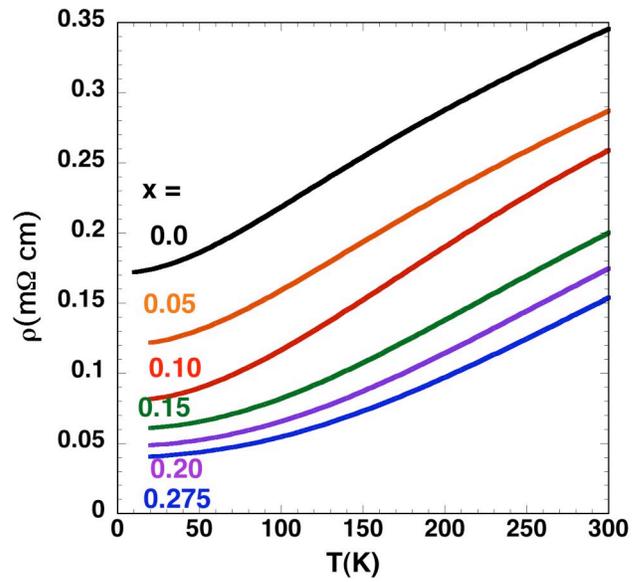

Fig. 4 Temperature dependence of the electrical resistivity from the model using Eq. 8 and the experimental findings of Ref. 21. $\rho_1= 2.4 \times 10^{20}$ (m$\Omega$ cm)(electrons/cm$^3$), $\rho_2=5.25 \times 10^{14}$ (m$\Omega$ cm K$^{-2}$)(carriers/cm$^3$). The model only applies to temperatures above $T_c$ and the structural/magnetic phase transitions.

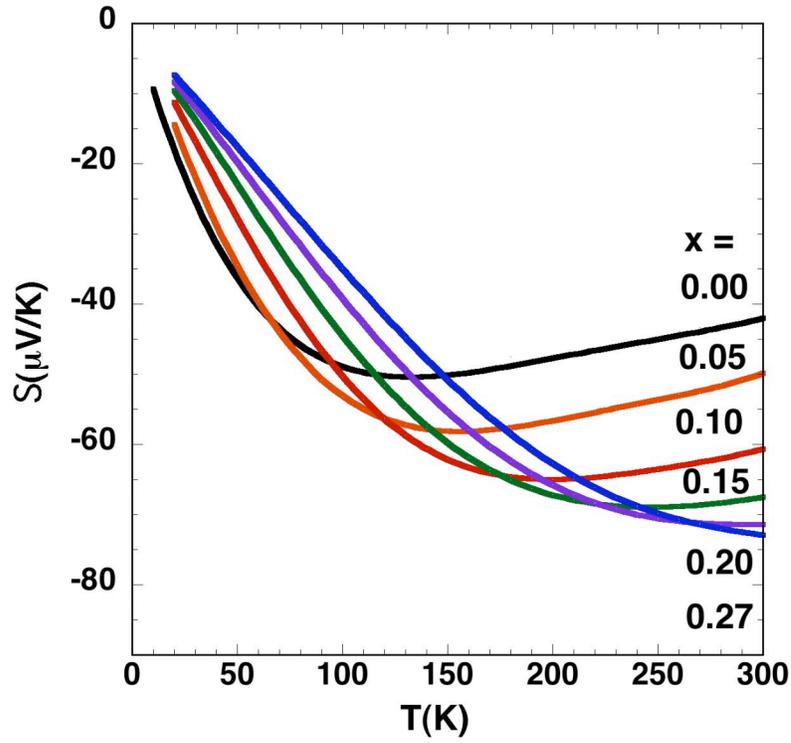

Fig. 5 Predicted variation of the Seebeck coefficient with temperature and Co doping x for Ba(Fe$_{1-x}$Co$_x$)$_2$As$_2$, for temperatures above T$_c$ and the magnetic/structural phase transition.

**References**

1. Y. Kamihara, T. Watanabe, M. Hirano and H. Hosono, J. Am. Chem. Soc. **130**, 3296 (2008).

2. G. F. Chen, Z. Li, D. Wu, G. Li, W. Z. Hu, J. Dong, P. Zheng, J. L. Luo and N. L. Wang, Phys. Rev. Lett. **100** 247002 (2008).

3. Z. Ren, J. Yang, W. Lu, W. Yi, G. Che, X. Dong, L. Sun and Z. Zhao, Materials Research Innovations **12**, 1 (2008).

4. Z. A. Ren, J. Yang, W. Lu, W. Yi, X. L. Shen, Z. C. Li, G. C. Che, X. L. Dong, L. L. Sun, F. Zhou and Z. X. Zhao Europhys. Lett. **82** 57002 (2008).

5. Z. Ren, W. Lu, J. Yang, W. Yi, X. Shen, Z. Li, G. Che, X. Dong, L. Sun, F. Zhou and Z. Zhao, Chin. Phys. Lett. **25**, 2215 (2008).

6. P. Cheng, L. Fang, H. Yang, X. Zhu, G. Mu, H. Luo, Z. Wang and H. Wen, Science in China G **51**, 719 (2008).

7. M. Rotter, M. Tefel, D. Johrendt, Phys. Rev. Lett. **101**, 107006 (2008).

8. X. C. Wang, Q. Q. Kiu, Y. X. Lv, W. B. Gao, L. X. Yang, R. C.Yu, F. Y. Li, C. Q. Jin, Solid State Comm. **148**, 538 (2008).

9. A. S. Sefat, M. A. McGuire, B. C. Sales, R. Y. Jin, J. Y. Howe and D. Mandrus, Phys. Rev. B. **77**, 174503.

10. M. Rotter, M. Tegel, I. Schellenberg, W. Hermes, R. Pottgen, and D. Johrendt, Phys. Rev. B. **78**, 020503(R) (2008).

11. M. A. McGuire et al., Phys. Rev. B. **78**, 094517, (2008).

12. A. S. Sefat, R. Y. Jin, M. A. McGuire, B. C. Sales, D. J. Singh, and D. Mandrus, Phys. Rev. Lett. **101**, 117004 (2008).

13. N. Ni, M. E. Tillman, J. –Q Yan, A. Kracher, S. T. Hannahs, S. L. Bud'ko, and P. C. Canfield, Phys. Rev. B. 214515 (2008).

14. F. Ning, K. Ahilan, T. Imai, A. S. Sefat, R. Y. Jin, M. A. McGuire, B. C. Sales, and D. Mandrus, J. Phys. Soc. Jpn. **78**, 013711 (2009).

15. R. Y. Jin et al. unpublished

16. J. H. Tapp, Z. Tang, B. Lv, K. Sasmal, B. Lorenz, P. C. W. Chu and M. Guloy, Phys.

Rev. B. **78**, 060505(R) (2008).

17. D. J. Singh and M. H. Du, Phys. Rev. Lett. **100**, 237003 (2008).

18. I. R. Shein and A. L. Ivanovskii, JETP Lett. **88**, 107 (2008)

19. D. J. Singh, Phys. Rev. B. **78**, 094511 (2008)

20. J-H Chu, J. G. Analytis, C. Kucharczyk, and I. R. Fisher, Phys. Rev. B. **79**, 014506 (2009).

21. F. Rullier-Albenque, D. Colson, A. Forget, H, Alloul, arXiv:0903.5243v1

22. Q. Huang, Y. Qui, W. Bao, M. A. Green, J. W. Lynn, Y. C. Gasparovic, T. Wu, G. Wu and X. H. Chen, Phys. Rev. Lett. **101**, 257003 (2008).

23. M. D. Lumsden et al., Phys. Rev. Lett, **102**, 107005 (2008).

24. P. Vilmercati, A. Fedorov, I. Vobornik, U. Manju, G. Panaccione, A. Goldoni, A. S. Sefat, M. A. McGuire, B. C. Sales, R. Jin, D. Mandrus, D. J. Singh and N. Mannella, Phys. Rev. B. **79**, 220503(R) (2009).

25. M. Yi et al. arXiv.0902.2628

26. Y. Yin, M. Zech, T. L. Williams, X. F. Wang, G. Wu, X. H. Chen, J. E. Hoffman, Phys. Rev. Lett. **102**, 097002 (2008).

27. S.M. Sze, "Physics of Semiconductors Devices" , (John Wiley and Sons, New York 1981) pp. 7-60.

28. H. J. Goldsmid, "Electronic Refrigeration", (Pion Limited, London, 1986) pp. 29-48.

29. S. Blundell, "Magnetism in Condensed Matter", (Oxford University Press, Oxford 2001) p. 144

30. C. Kittel, "Introduction to Solid State Physics, 7$^{th}$ edition", (John Wiley and Sons, Hoboken, New Jersey, 1996) p. 232.

31. B. C. Sales, D. Mandrus, B. C. Chakoumakos, V. Keppens, and J. R. Thompson, Phys. Rev. B. **56**, 15081 (1997).

32. N. W. Ashcroft and N. D. Mermin, "Solid State Physics", (Holt, Rinehart, and Winston, New York 1976) p. 348.

33. L. Fang et al. arXiv:0903.2418

34. E. D. Mun, S. L. Bud'ko, N. Ni, P. C. Canfield arXv:0906.1548v1